\documentclass[twocolumn,nofootinbib,preprintnumbers,amsmath,amssymb]{revtex4}

\usepackage{graphicx}
\usepackage{dcolumn}
\usepackage{bm}
\usepackage{amsmath}

\def \be {\begin{equation}}
\def \ee {\end{equation}}

\begin{document}

\title{How Pulsars Shine: Poynting Flux Annihilation}

\author{Andrei Gruzinov}

\affiliation{ CCPP, Physics Department, New York University, 4 Washington Place, New York, NY 10003
}

\begin{abstract}

Recent solution of a pulsar (although likely correct, as agreeing with phenomenology of weak pulsars without a single adjustable) is somewhat unsatisfactory on account of being heavily numerical. Here the main features of a pulsar -- (i) the existence of a force-free zone, (ii) bounded by a non-force-free radiation zone, (iii) where colliding  Poynting fluxes annihilate into curvature radiation -- are reproduced as an exact solution of the same pulsar equations but in a much simpler geometry. 

\end{abstract}

\maketitle

\section{Introduction}

We first give a qualitative description of a pulsar according to \cite{Gruzinov2013}, \S\ref{pic}. This description is based on numerical solutions of Aristotelian Electrodynamics (AE = Electrodynamics of Massless Charges). AE equations are summarized in \S\ref{AE}. In \S\ref{dev} we give an exact solution of AE equations for ``the Device'' which was specially chosen to be exactly solvable and similar to real pulsars. Namely, the Device features: (i) a force-free zone, (ii) bounded by a non-force-free radiation zone, (iii) where colliding  Poynting fluxes annihilate into curvature radiation. 

\section{Pulsar according to \cite{Gruzinov2013}}\label{pic}

\subsection{Magnetosphere}

For clarity, consider an axisymmetric pulsar. As shown in Fig.1, the star is surrounded by a Force-Free Zone {\bf I}, where energy is not dissipated. Part of the Force-Free Zone -- the Corotation Zone {\bf I'} -- is characterized by a purely toroidal Poynting flux. In the Corotation Zone the ExB drift just rotates the charges at the angular velocity of the star.

In the non-corotating part of the Force-Free Zone, the Poynting flux has an outward poloidal component. Most of the Poynting flux flowing in the Force-Free Zone ultimately enters a non-force-free Radiation Zone {\bf II}, where the Poynting flux is damped into curvature radiation. 

\begin{figure}[bth]
  \centering
  \includegraphics[scale=0.75, trim = 6.5cm 11cm 5.5cm 8cm, clip]{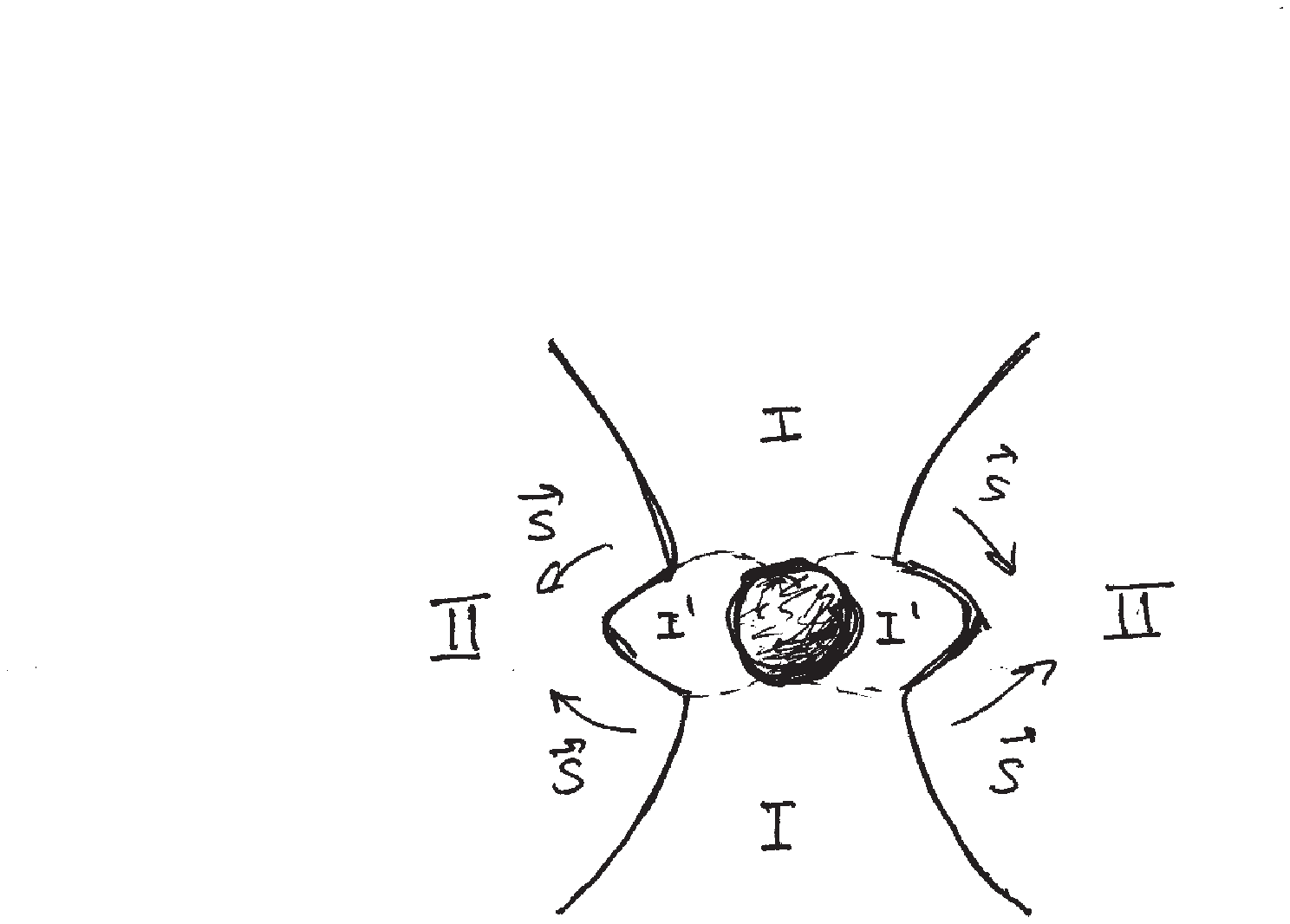}
\caption{{\bf I} -- Force-Free Zone; {\bf I'} -- Corotation Zone; {\bf II} -- Radiation Zone, {\bf S} -- colliding Poynting fluxes.} \label{bol}
\end{figure}

\subsection{Weak and Non-Weak Pulsars}

In {\it weak} pulsars {\it substantial} plasma production occurs only in the Force-Free Zone (charges are created by avalanches and pulled out of the star \cite{Ruderman1975}). {\it Substantial} means comparable to Goldreich-Julian density \cite{Goldreich1969} per rotation. Weak pulsars are the easiest to calculate, because one may postulate an arbitrary rate of plasma production in the Force-Free Zone (as long as the plasma production rate is postulated to be high for large proper electric fields). Only weak pulsars have been calculated by \cite{Gruzinov2013}. This calculation reproduces so much of the weak pulsar phenomenology as seen by Fermi \cite{Fermi2013} that it must be correct or very close to correct.

{\it Non-weak} pulsars are pulsars which are not {\it weak}. In non-weak pulsars, photon-photon collisions give substantial pair production in the Radiation Zone. Now the true rate of pair production becomes relevant, and must be self-consistently included into the pulsar calculation. This appears doable, but it has not been done yet.

\subsection{Emission of Weak Pulsars}

The calculation of \cite{Gruzinov2013} gives not only the electromagnetic field but also the electron and positron charge densities everywhere in the Radiation Zone. Knowing the fields and the particle densities one calculates the resulting (curvature) radiation.

\section{Aristotelian Electrodynamics}\label{AE}

The key to solving the pulsar is AE: positive and negative charges move at the speed of light
\be\label{ae}
{\bf v}_{\pm}={{\bf E}\times {\bf B}\pm(B_0{\bf B}+E_0{\bf E})\over B^2+E_0^2},
\ee
emitting (along ${\bf v}_{\pm}$) curvature power (per charge)
\be 
q=ecE_0,
\ee
with synchrotron spectrum of critical photon energy 
\be\label{ecrit}
E_{\rm c}=(3/2)^{7/4}c\hbar e^{-3/4}E_0^{3/4}K^{-1/2}.
\ee

In the above: $E_0$ is the proper electric field scalar and $B_0$ is the proper magnetic field pseudoscalar defined by 
\be\label{prop}
B_0^2-E_0^2=B^2-E^2,~ B_0E_0={\bf B}\cdot {\bf E},~ E_0\geq 0;
\ee
$K$ is the curvature of the particle trajectory
\be
K=|(\partial _t+{\bf v}_\pm \cdot \nabla ) {\bf v}_\pm |.
\ee

AE is valid under the assumption of strong radiation over-damping. To the best of my knowledge, AE was first derived by \cite{Finkbeiner1989}. The easiest way to get eq.(\ref{ae}) is to construct a Lorentz-covariant expression for a future-directed null 4-vector solely in terms of the electromagnetic field tensor -- this then must be the direction of the 4-momentum of an over-damped ultra-relativistic charge. All other expressions follow trivially. 

Knowing how the charges move and radiate and postulating (for weak pulsars) or self-consistently calculating (for non-weak pulsars) the pair production rate, one solves the pulsar.

\section{The Device}\label{dev}

Here we 
\begin{itemize}
\item give a two-dimensional reduction of AE, \S\ref{t1};
\item derive equations of stationary 2D AE, \S\ref{t2};
\item describe the Device mimicking the pulsar, \S\ref{t3};
\item solve the equations of stationary 2D AE for the Device boundary conditions, \S\ref{t4}; 
\item discuss the physics of the solution, \S\ref{t5}.
\end{itemize}

\subsection{Electron AE in two dimensions}\label{t1}

Assume that the electromagnetic field is two dimensional in the following sense:
\be
{\bf E}=(E_x,E_y,0),~~{\bf B}=(0,0,B), ~~\partial _z=0.
\ee
Further assume that electrons are the only charges present. Then the basic AE equation (\ref{ae}) gives the following Ohm's law
\be\label{ohm}
{\bf j}=-\left( {B\hat{z}\times {\bf E}+E_0{\bf E}\over B^2+E_0^2} \right)\nabla \cdot {\bf E},
\ee
where 
\be
E_0^2=(E^2-B^2)\theta (E^2-B^2).
\ee

The Ohm's law is used to solve Maxwell equations
\be\label{2dae}
\dot{B}=-\hat{z}\cdot \nabla \times {\bf E},~~\dot{\bf E}=-\hat{z}\times \nabla B-{\bf j}.
\ee

Once the {\bf E} and B fields are known, one calculates the resulting emission using the formulas of \S\ref{AE} (recalling that the electron number density is $-\nabla \cdot {\bf E}/e$) .

\subsection{Stationary Electron AE in two dimensions}\label{t2}

We will not study the transient process, but only the steady state, $\partial _t=0$. Then equations (\ref{2dae}, \ref{ohm}) give
\be
{\bf E}=-\nabla \phi,
\ee
\be\label{2daes}
\hat{z}\times \nabla B=\left( {B\hat{z}\times \nabla \phi+E_0\nabla \phi\over B^2+E_0^2} \right)\nabla ^2\phi.
\ee
To write Eq.(\ref{2daes}) in a scalar form, we multiply both sides on $\nabla \phi$ and on $\hat{z} \times \nabla \phi$:

\be\label{s1}
{B(\nabla \phi)^2\over B^2+E_0^2}\nabla ^2\phi =\nabla \phi \cdot \nabla B,
\ee
\be\label{s2}
{E_0(\nabla \phi)^2\over B^2+E_0^2}\nabla ^2\phi =-\hat{z}\cdot (\nabla \phi \times \nabla B).
\ee

The field is force-free if $B^2>(\nabla \phi)^2$. Then we have $E_0=0$ and eq.(\ref{s2}) gives 
\be\label{gs1}
B=B(\phi).
\ee
Now from eq.(\ref{s1}) we get, as we should, the ``Grad-Shafranov'' equation (Scharlemant-Wagoner \cite{Scharlemant1973} in the pulsar context):
\be\label{gs2}
\nabla ^2 \phi = B{dB\over d\phi}.
\ee

The field is non-force-free if $B^2<(\nabla \phi)^2$. Then $E_0^2=(\nabla \phi)^2-B^2$ and eqs.(\ref{s1}, \ref{s2}) read
\be\label{s11}
B\nabla ^2\phi =\nabla \phi \cdot \nabla B,
\ee
\be\label{s21}
E_0\nabla ^2\phi =-\hat{z}\cdot (\nabla \phi \times \nabla B).
\ee

\subsection{The Device}\label{t3}

Motivated by Fig.1, where Poynting fluxes collide at the equatorial plane, we consider the following device.

\begin{figure}[bth]
  \centering
  \includegraphics[scale=0.4, trim= 1cm 14cm 0 -1.5cm, clip]{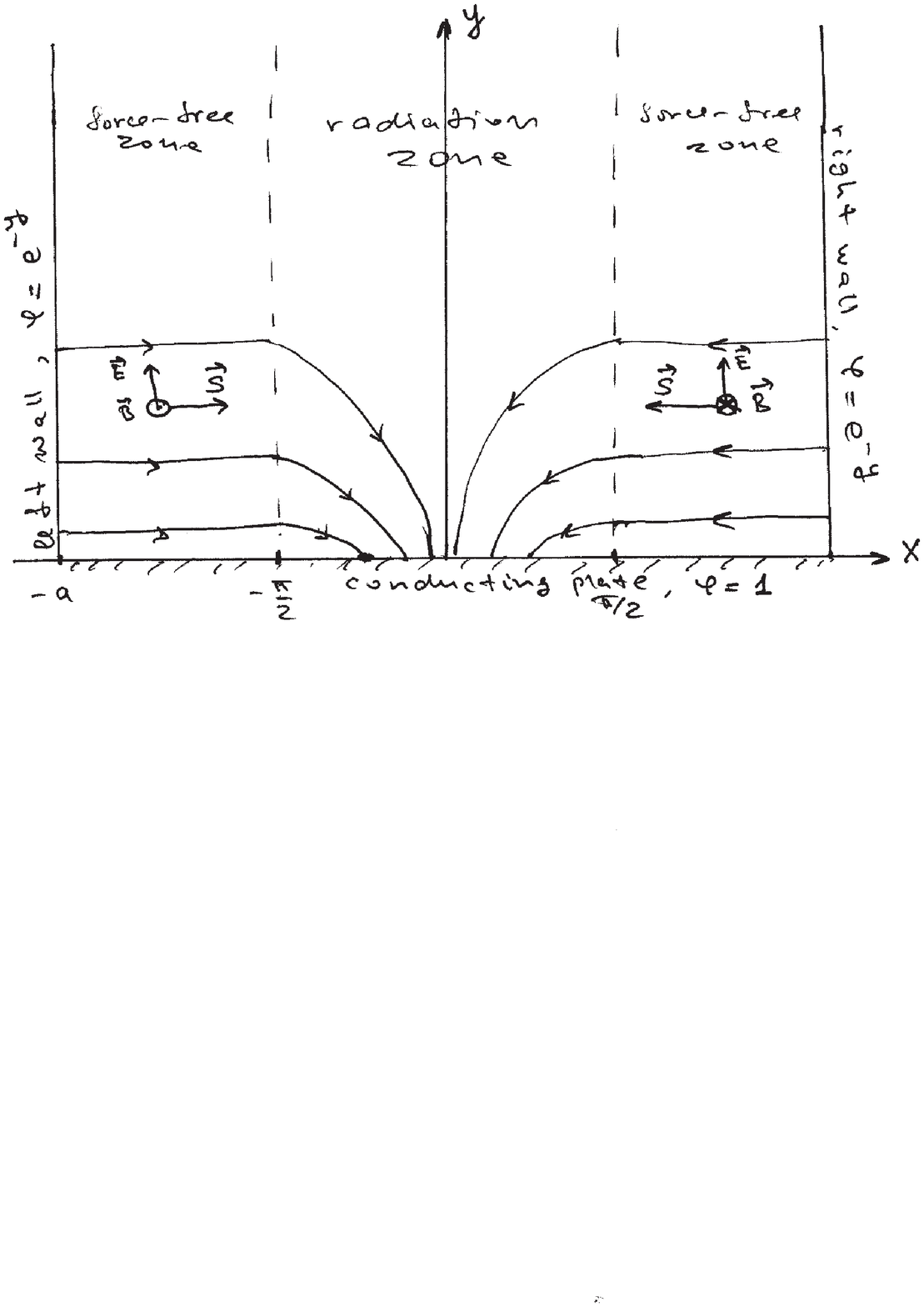}
\caption{The Device.} \label{f2}
\end{figure}

As shown in Fig.2., there is a conducting plate at the x-axis extending from $x=-a$ to $x=a$, with arbitrary $a>\pi/2$. The plate is kept at a fixed potential $\phi=1$. The plate does not emit electrons, but absorbs all electrons reaching it. There are also two walls perpendicular to the plate, running from $y=0$ to $y=\infty$ along $|x|=a$. The walls are kept at fixed potentials $\phi=e^{-y}$. So long as the field right outside the wall is electric like, $E>|B|$, with a negative normal component of the electric field, the electrons are very copiously pulled out. Here `` very copiously'' should be understood as follows -- the device saturates only at a zero normal electric field (or at a null-like, $E=|B|$, field) and in the saturated state some electrons can still be pulled out at a rate needed to sustain the saturation.

We start off without any charges inside the Device, with $B=0$, and with potential electric field ${\bf E}=-\nabla \phi$ which is given by $\nabla ^2\phi =0$ and the boundary values of $\phi$.  The walls start to emit electrons creating $B$ and changing ${\bf E}$, as described by Maxwell equations (\ref{2dae}) (with appropriate boundary conditions at the walls describing the electron pull-out rate).

\subsection{Exact Solution}\label{t4}

Assume that the Device does saturate. The resulting fields should satisfy stationary AE equations of \S\ref{t2}. One can check that the following expressions do solve the equations and satisfy the boundary conditions specified above:
\be
\phi (x,y)=e^{-y},
\ee
\be 
B(x,y) = \left\{ \begin{array}{rl}
-e^{-y}sign(x), & ~~~|x|>{\pi \over 2}; \\
-e^{-y}\sin (x), & ~~~|x|<{\pi \over 2}. \\
\end{array}\right.
\ee

The domain ${\pi \over 2}<|x|<a$ is a force-free zone. Here the Grad-Shafranov equations (\ref{gs1}, \ref{gs2}) are obeyed:
\be
E_0=0,~~B=\pm \phi,~~ \nabla ^2\phi=\phi=B{dB\over d\phi}.
\ee

The domain $|x|<{\pi \over 2}$ is a radiation zone. Here the non-force-free stationary AE equations (\ref{s11}, \ref{s21}) are obeyed
\be
E_0=e^{-y}\cos (x), 
\ee
\be
\nabla \phi \cdot \nabla B=\partial _y\phi \partial _yB=\phi B=B\nabla^2\phi,
\ee
\be
\hat{z}\cdot \nabla \phi \times \nabla B=-\partial _y\phi \partial _xB=-E_0\nabla ^2\phi .
\ee

\subsection{Discussion}\label{t5}

The electromagnetic field and the corresponding Poynting fluxes in the force-free zones are shown in Fig.(2). In the left force-free zone electrons are moving to the right (along the lines of constant $B$ shown in the figure). In the right force-free zone electrons are moving to the left.

The two electron beams carrying the Poynting fluxes are so polarized (opposite signs of $B$ with collinear ${\bf E}$), that if they were to interpenetrate, a purely-electric region would have been created, where the electrons would have to move strictly down, rather than towards each other. A radiation zone must therefore separate the two force-free zones.

In the radiation zone, the Poynting flux remains parallel to the x-axis, flowing towards the y-axis, but none of it reaches the y-axis: ${\bf S}=(E_yB, -E_xB, 0)=(-e^{-2y}\sin(x),0,0)$. The entire Poynting flux gets annihilated into curvature radiation. 

To discuss radiation, it makes sense to restore dimensions. We can do it by prescribing the wall potential as $\phi =FRe^{-y/R}$ where $F$ has dimensions of electromagnetic field and $R$ has dimensions of length. Then the curvature of electron trajectories in the radiation zone is $\sim R^{-1}$, the proper electric field is  $\sim F$. According to the formulas of \S\ref{AE}, the Device emits curvature photons of characteristic energy 
\be
E_{\rm c}\sim c\hbar e^{-3/4}F^{3/4}R^{1/2}.
\ee

Calling $L\sim cF^2R^2$ the power, erg/s, we get (as we should) the formula \cite{Gruzinov2013} expressing the photon cutoff energy $E_{\rm c}$ in terms of the power $L$ and the size of the emitting region $R$:
\be
E_{\rm c}\sim {mc^2\over \alpha}{\rm Ar}^{3/8},
\ee
where $m$ is the mass of electron, $\alpha$ is the fine structure constant,
\be 
{\rm Ar}\equiv {L\over L_e}\left( {R\over r_e}\right) ^{-2/3},
\ee
is the Aristotle number of the Device, $r_e={e^2\over mc^2}=2.8\times 10^{-13}$cm is the classical electron radius and $L_e={mc^3\over r_e}=8.7\times 10^{16}$erg/s is the ``classical electron luminosity''. AE is applicable at large Aristotle numbers, 
\be 
{\rm Ar}\gg 1. 
\ee

\section{Conclusion}

\begin{itemize}
\item Pulsars shine by annihilating colliding Poynting Fluxes into curvature radiation.
\item To solve a pulsar one can use AE.
\item The Device, being exactly solvable in the AE weak-pulsar limit, can prove useful in testing  numerical codes, especially when the weak-pulsar and/or AE approximations are lifted.
\end{itemize}


\begin{thebibliography}{99}

\bibitem{Gruzinov2013}
A. Gruzinov, arXiv:1309.6974, 1310.1894, 1310.3261, 1310.5382 (2013)

\bibitem{Ruderman1975}
 M. A. Ruderman, P. G. Sutherland, Astrophys.\ J.\ {\bf 196}, 51 (1975)

\bibitem{Goldreich1969}
 P. Goldreich, W. H. Julian, Astrophys.\ J.\ {\bf 157}, 869 (1969)

\bibitem{Fermi2013}
The Fermi-LAT collaboration, arXiv:1305.4385 (2013)

\bibitem{Scharlemant1973}
E. T. Scharlemant, R. V. Wagoner, Astrophys.\ J.\ {\bf 182}, 951 (1973)


\bibitem{Finkbeiner1989}
B. Finkbeiner, H. Herold, T. Ertl, H. Ruder, Astron. Astrophys. {\bf 225}, 479 (1989)



\end{thebibliography}
\end{document}